\documentclass[twocolumn,reprint,preprintnumbers,amsmath,amssymb,aps,prl,nofootinbib,superscriptaddress ]{revtex4-1}

\usepackage{graphicx}
\usepackage{dcolumn}
\usepackage{bm}
\usepackage{epsfig,psfrag,graphics,verbatim}
\usepackage{xcolor}
\usepackage{float}
\usepackage{amsmath,amssymb}
\usepackage{siunitx}
\usepackage{hyperref}
\usepackage{xfrac}
\usepackage{slashed}
\usepackage{ulem}

\hypersetup{
  colorlinks=true,
  linkcolor=blue,
  citecolor=blue,
  urlcolor=blue,
  hyperfootnotes=true
}

\def\hhref#1{\href{http://arxiv.org/abs/#1}{#1}} 

\newcommand{\beq}{\begin{eqnarray}}
\newcommand{\eeq}{\end{eqnarray}}

\newcommand{\gsim}{\lower.7ex\hbox{$\;\stackrel{\textstyle>}{\sim}\;$}}
\newcommand{\lsim}{\lower.7ex\hbox{$\;\stackrel{\textstyle<}{\sim}\;$}}

\newcommand\GeV{\rm GeV}
\newcommand\MeV{\rm MeV}
\newcommand\cm{\rm cm}

\newcommand{\be}{\begin{equation}} 
\newcommand{\ee}{\end{equation}}
\newcommand{\bea}{\begin{equation}\begin{aligned}}
\newcommand{\eea}{\end{aligned}\end{equation}}
\newcommand{\td}{{\rm d}}

\def\circa#1{\,\raise.3ex\hbox{$#1$\kern-.75em\lower1ex\hbox{$\sim$}}\,}

\begin{document}

\preprint{CERN-TH-2020-165}
\preprint{TTP20-039}
\preprint{P3H-20-070} 

\title{{\sc Xenon1T} excess from electron recoils of non-relativistic Dark Matter}

\author{Dario Buttazzo}
\affiliation{INFN, Sezione di Pisa, Largo Bruno Pontecorvo 3, I-56127 Pisa, Italy}

\author{Paolo Panci}
\affiliation{Dipartimento di Fisica ``E. Fermi'', Universit{\`a} di Pisa, Largo Bruno Pontecorvo 3, I-56127 Pisa, Italy}
\affiliation{INFN, Sezione di Pisa, Largo Bruno Pontecorvo 3, I-56127 Pisa, Italy}

\author{Daniele Teresi}
\affiliation{CERN, Theoretical Physics Department, 1211 Geneva 23, Switzerland}
\affiliation{Dipartimento di Fisica ``E. Fermi'', Universit{\`a} di Pisa, Largo Bruno Pontecorvo 3, I-56127 Pisa, Italy}

\author{Robert Ziegler}
\affiliation{Institut f\"ur Theoretische Teilchenphysik,  Karlsruhe Institute of Technology,  Karlsruhe,  Germany}

\date{\today}

\begin{abstract}
We show that electron recoils induced by non-relativistic Dark Matter interactions can fit well the recently reported Xenon1T excess, if they are mediated by a light pseudo-scalar in the MeV range.
 This is due to the favorable momentum-dependence of the resulting scattering rate, which partially compensates the unfavorable  kinematics that tends to strongly suppress keV electron recoils. We study the phenomenology of the mediator and identify the allowed parameter space of the  {\sc Xenon1T} excess which is compatible with all 
experimental limits. We also find that the anomalous magnetic moments {$(g-2)_{\mu,e}$} of muons  and electrons can be simultaneously explained in this scenario, at 
the price of  a fine-tuning in the couplings of the order of a few percent.

\end{abstract}

\maketitle

\section{Introduction}

Recently, the {\sc Xenon1T} collaboration has announced the results of a search for Dark Matter (DM) using electronic recoils with a 0.65 ton/years of exposure. An unexpected peak of electronic recoil events over  the nominal background has been reported~\cite{Aprile:2020tmw}. The excess corresponds to 53 events in the 1$-$7 keV energy window, mainly located in the energy bins close to the experimental threshold. 

Several possibilities for the origin of this signal have been proposed. The {\sc Xenon1T}  collaboration itself analyzed the signal in terms of solar axion absorption or solar neutrinos scattering off electrons with an enhanced magnetic moment. While these interpretations have the  advantage of not suffering from a look-elsewhere effect (LEE), essentially because their scale is fixed by the Sun temperature, they are strongly disfavored by astrophysical bounds~\cite{DiLuzio:2020jjp, Gao:2020wer}. Another option 
is scattering due to a fast component of DM~\cite{Kannike:2020agf}, which however requires non-trivial model-building (see e.g. Ref.~\cite{McKeen:2020vpf}). Absorption of bosonic keV-scale DM (see e.g. Ref.~\cite{Takahashi:2020bpq}) or, in general, models where the keV scale is determined by kinematic features (see e.g. Ref.~\cite{Smirnov:2020zwf})  suffer of LEE and thus lower their statistical preference with respect to the Standard Model. 

In this letter, we show that the excess can be explained by {\it standard} electron recoils of GeV or heavier DM, as long as the DM-$e$ interactions are mediated by a {\it pseudo-scalar} particle. The main challenge in explaining the excess by scattering~\cite{Bloch:2020uzh} is to get
a signal in the 2--4 keV bins and yet be compatible with bounds at lower recoil energies where a significant excess is not seen, even taking into account the suppressed detector sensitivity. While the scattering kinematics of non-relativistic DM tends to strongly suppress  keV recoils (which are possible only in the momentum-distribution tails of the  xenon atomic wave-functions), the interaction mediated by a pseudo-scalar increases with the exchanged momentum
, partially compensating the unfavorable kinematics and allowing for a good fit of the excess.
It is worth stressing  that since the signal is due to the tail of the electron atomic distribution, our explanation does not suffer of LEE.
{Indeed, our model simply predicts a signal continuously decreasing with energy in the {\sc Xenon1T} region, so that the largest effect is always close to the experimental threshold, which is indeed the case of the excess. Signals that peak away from the threshold would not be explained by our model.}

\section{KeV electron recoils from pseudoscalar mediator}
We consider a simplified model with a pseudo-Nambu-Goldstone boson $a$ that couples derivatively to electrons and photons, as well as to a Dirac fermion $\chi$ that will account for DM. The relevant interaction Lagrangian is given by
\be\label{eq:Lbase2}
\mathcal L =  \frac{\partial_\mu a}{\Lambda}  \left( c_{\chi a} \bar \chi \gamma^\mu\gamma_5 \chi +   c_{e a} \bar e \gamma^\mu\gamma_5 e \right) + \frac{\alpha}{2 \pi} \, C_{\gamma \gamma}  \frac{a}{\Lambda} F \tilde F  \, , 
\ee
where $F \tilde F \equiv 1/2 \, \epsilon^{\mu \nu \rho \sigma} F_{\mu\nu}  F_{\rho \sigma}$. 
 For later purposes, it will 
 be more convenient to work with  the following Lagrangian
\be\label{eq:Lbase1}
\mathcal L =  - i  a \left( g_\chi \bar \chi \gamma_5 \chi + g_e  \bar e \gamma_5 e \right) + \frac{\alpha}{2 \pi} \, \widetilde C_{\gamma \gamma}  \frac{a}{\Lambda} F \tilde F  \, , 
\ee
which is equivalent to Eq.~\eqref{eq:Lbase2} if the effective couplings of $a$ to fermions are $g_i \equiv 2m_i c_{ia}/\Lambda$, $i = e, \chi$, and $\widetilde C_{\gamma\gamma} \equiv C_{\gamma\gamma} + c_{ea}$. Upper bound on these couplings are obtained from perturbative unitarity, by requiring that partial waves of total angular momentum $J = 0$ are smaller than 1/2, giving $g_i < \sqrt{8\pi/3}$ \cite{Cornella:2019uxs}.

The amplitude for $\chi\,e^-\to \chi\,e^-$ scattering is
\be \label{eq:ampl}
\mathcal A =  \, \bar \chi \gamma_5 \chi \,  \frac{ g_{\chi}  g_e}{q^2+m_a^2}\bar e \gamma_5 e \ ,
\ee
where $q \equiv \left| {\bf q} \right| $ is the size
of the three-momentum transferred in the scattering process, which typically is of the order of few MeV. 
Following the notation of Ref.~\cite{Roberts:2019chv}, the  velocity-averaged differential cross-section is given by
\begin{equation}\label{eq:sigmav}
	\frac{\td \langle \sigma v \rangle}{\td E_R} = \frac{\bar \sigma_e}{2 m_e} \,\int d v \, \frac{f(v)}{v} \! \int_{q_-}^{q_+}\!\!\! q \, \td q |F_\chi(q)|^2 \, \frac{4a_0^2}{\alpha^2}   K_5(E_R,q) ,
\end{equation}
where $a_0 = 1/(\alpha m_e)$ is the Bohr radius. The limits of integration for the exchanged momentum are
$
	q_\pm = m_{\chi} v \pm \sqrt{m_{\chi}^2 v^2 - 2 m_\chi E_R} ,
$
with $E_R$ the electron recoil energy, and $f(v)$ is the DM distribution in the Earth frame normalized as $\int dv f(v) = 1$. We use a truncated Maxwell-Boltzmann distribution with mean velocity of 220 km/s, average Earth's velocity of 240 km/s and  galactic escape velocity of 544 km/s. 
We have normalized the cross-section in Eq.~\eqref{eq:sigmav} by using the reference {\it contact} cross-section for DM scattering on {\it free} electrons at $q=\alpha m_e$,
\begin{equation}
	\bar \sigma_e = \frac{m_e^2}{16 \pi} \frac{g_\chi^2 g_e^2}{m_a^4} \frac{q^4}{m_\chi^2 m_e^2}\bigg|_{q = \alpha m_e} \, .
\end{equation}

$F_\chi(q)$ is the form factor that includes the contribution of the propagator and 
the DM pseudo-scalar vertex to the  amplitude in Eq.~\eqref{eq:ampl},
\be \label{eq:F}
F_\chi(q) = \frac{q}{\alpha m_e}\frac{m_a^2}{q^2 + m_{a}^2} \ .
\ee

The  contribution  of the electron pseudo-scalar vertex is instead included in the pseudo-scalar atomic ionization function $K_5(E_R,q)$. This is the squared matrix element of the pseudo-scalar electron current in Eq.~\eqref{eq:ampl} between free and atomic states and contains relativistic corrections that are relevant for $q\sim$ MeV. In the non-relativistic limit the ratio between the pseudo-scalar and scalar ionization functions is $K_5(E_R,q)/K(E_R,q) \propto  (q/2m_e)^2$ due to the different Lorentz structure of the two electron currents. For $E_R \sim $ keV, $K_5$ is dominated by the $3s$ and $4s$ orbitals, the former starting at $E_R>1.17$ keV. We use the relativistic $3s$ pseudo-scalar ionization factor  provided in Ref.~\cite{Roberts:2016xfw}\footnote{Below the $3s$ threshold, we approximate the (small) $4s$ ionization factor as having the same momentum dependence as the 3s one, with the overall factor determined by the scalar ones at $q \approx \rm MeV$.}. It is worth noticing that since the pesudo-scalar and scalar atomic  ionization functions are similar for $q\approx$ MeV (see right panel of Fig.~26 in Ref.~\cite{Roberts:2016xfw}), the factor $4/\alpha^2$ in~\eqref{eq:sigmav} is needed because we are normalizing $\bar \sigma_e$ at $q=\alpha m_e$.  Indeed for $q=\alpha m_e$, $K_5(E_R,q)/K(E_R,q) \propto  (\alpha/2)^2$ which is exactly the suppression one gets between the normalized cross sections of the pure ($\bar \chi \gamma_5 \chi \,  \bar e \gamma_5 e$) and CP-violating  (i.e.~$\bar \chi \gamma_5 \chi \,  \bar e  e$) pseudo-scalar interactions. On the other hand for $q\sim$ MeV, the DM-electron collisions are relativistic and therefore the differential cross sections of the two interactions must be of the same order as one can check from Eq.~\eqref{eq:sigmav}.

The differential scattering rate is given by
$
\td R/\td E_R = N_T n_\chi \, \td \langle \sigma v \rangle/\td E_R \,,
$
where $N_T \simeq 4.2 \times 10^{27} / {\rm ton}$ is the number  of Xe atoms per ton of detector, and $m_\chi n_\chi \simeq 0.4 \, \GeV /\cm^3$ is the local DM energy density. To compare our recoil spectra with the {\sc Xenon1T} results, we apply a Gaussian smearing with a detector resolution $\sigma_{\rm det} = 0.45 \, \mathrm{keV}$ \cite{Aprile:2020yad}, multiply by the efficiency given in Ref.~\cite{Aprile:2020tmw} and bin the data as in the spectrum given by the {\sc Xenon} collaboration.

\begin{figure}
$$\includegraphics[height=0.4\textwidth]{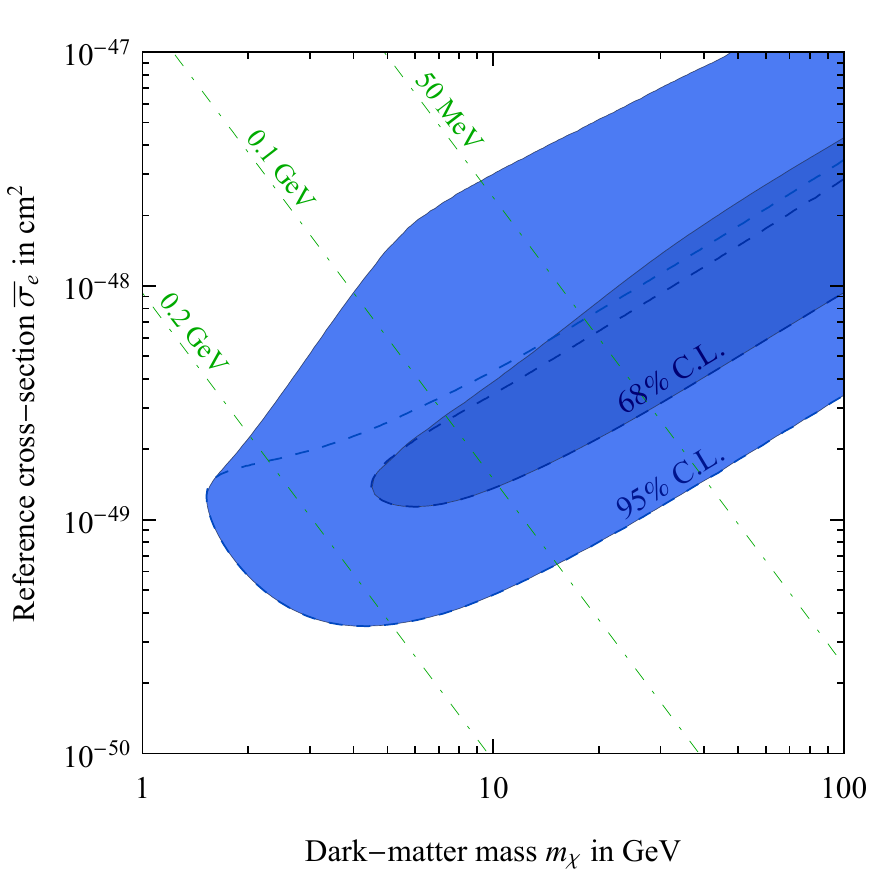}$$
\caption{Parameter space for reproducing the {\sc Xenon1T} excess as function of the DM mass and reference free electron cross-section, profiling  over the mass $m_a$ of the ALP mediator. We also show the results in the limit of contact interaction (dashed lines) as well as the corresponding interaction scale $m_a/(g_e g_\chi)^{1/2}$ (green dot-dashed lines).  \label{fig:sigma}}
\end{figure}

We perform a profile likelihood ratio fit, fixing the background to the best-fit spectrum given in Ref.~\cite{Aprile:2020tmw}. We have checked that including the overall magnitude of the background and the efficiency as nuisance parameters, the results are not significantly affected. Instead, as expected, including the possibility of a tritium background component with free amplitude in the fit decreases the significance of the excess and, as a consequence,  extends drastically the parameter space. We show the results in Fig.~\ref{fig:sigma} as a function of the DM mass and reference free electron cross-section. We present the results obtained both profiling  over the mediator mass $m_a$ and in the contact-interaction limit\footnote{Our results differ from the ones of the arXiv v2  of Ref.~\cite{Bloch:2020uzh}, that work in the contact-interaction limit. The discrepancy is due to the fact that in Ref.~\cite{Bloch:2020uzh} the pseudo-scalar interaction is treated as an effective factor $(q/2m_e)^2$ multiplying the {\it scalar} ionization factor, but this non-relativistic approximation is not valid for $q \gtrsim \MeV$.} 
$m_a \gg q \sim {\rm MeV}$. As could have been guessed, in a large region of parameter space the fit prefers a contact interaction, since this yields a spectrum that decreases slower with $q$ (see Eq.~\eqref{eq:F}, combined with $K_5 \sim 1/q^6$ for $q \gg {\rm MeV}$). Furthermore, it is worth stressing that due to the large non-relativistic suppression of the cross section the stringent bounds obtained from experiments that are looking for lower electron recoil energy (e.g.~the {\sc Xenon1T}  S2-only analysis~\cite{Aprile:2019xxb}) do not apply. Notice, however, that the required interaction scales are rather low, therefore we pass to study in detail the phenomenology of the ALP mediator.
\begin{figure}
$$\includegraphics[height=0.4\textwidth]{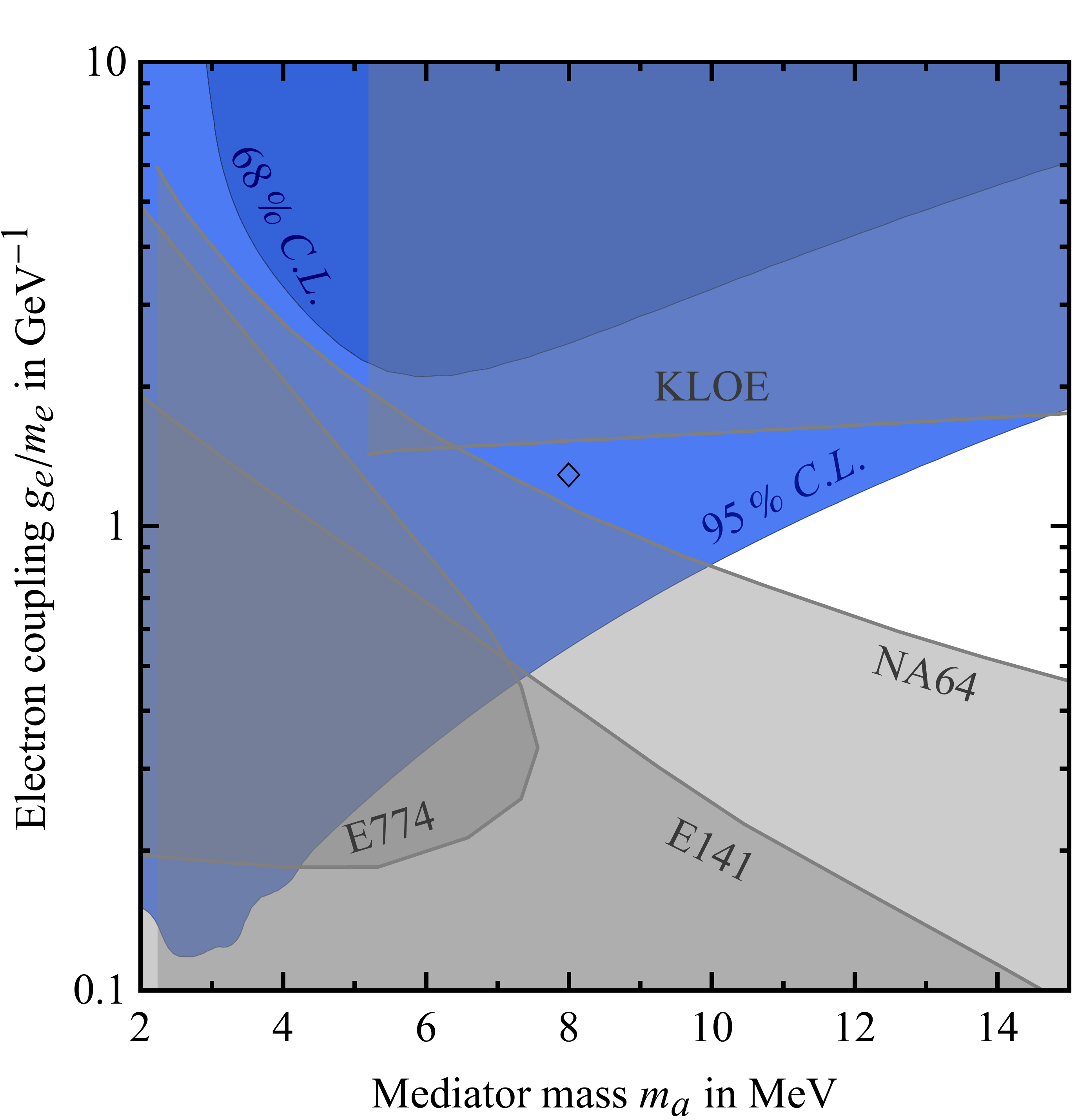}$$
\caption{Parameter space for the ALP $a$ with its DM coupling $g_\chi$ set at the partial-wave unitarity bound $g_\chi = \sqrt{8\pi/3}$, profiled over the DM mass. Exclusion limits at $90 \%$ C.L. from collider searches are also shown. The diamond denotes the benchmark point yielding the electron recoil spectrum at {\sc Xenon1T} shown in Fig.~\ref{fig:spectrum}. \label{fig:pheno}}
\end{figure}
\section{Collider bounds}
Several experiments have searched for light particles, and pose stringent limit on their couplings to leptons. Here we briefly recall the main experimental constraints that apply to our model.

The KLOE experiment has searched for a light new particle $a$ produced in association with a photon in $e^+e^-$ collisions, 
$e^+e^- \to \gamma a$, looking for the  prompt decay of $a$ into $e^+e^-$. While the original search was optimized for a massless vector particle, it has been recast for the case of a pseudo-scalar in Ref.~\cite{Alves:2017avw}. For ALP masses above 5 MeV, the KLOE result constrains the coupling to electrons to be smaller than roughly $10^{-3}$ {(smaller ALP masses cannot be probed due to the large irreducible background from radiative Bhabha scattering)}.

For lighter ALP masses or smaller couplings, the most stringent constraints come from electron beam-dump experiments at Fermilab (E774~\cite{E774}), SLAC (E141~\cite{E141}) and CERN (NA64~\cite{NA64}), searching for $e^+e^-$ decays of a short-lived particle produced from an electron beam stopped in an absorbing target. While E774 and E141 provide constraints directly on a pseudo-scalar boson, the results by NA64 are formulated as constraints on the kinetic mixing $\epsilon$ of a massive vector.  In order to recast the NA64 bound in terms of pseudo-scalar couplings, we use the simple approximate relation $g_e = \epsilon \sqrt{4 \pi \alpha}$, see e.g. Refs.~\cite{Bjorken:2009mm, Alves:2017avw} (a more refined recast could be performed along the lines of Ref.~\cite{Ema:2020fit}). Beam dump experiments with longer shielding, such as E137 at SLAC~\cite{E137} do not provide relevant constraints because here we are interested in very short lifetimes. 
Finally we note that photo-production and decay are always subleading with respect to electron production and decay for the relevant ALP mass range.

In Fig.~\ref{fig:pheno} we show the main collider and beam-dump constraints as grey regions in the $(m_a$--$g_e/m_e)$ plane by fixing the coupling of $a$ to DM to its  bound from perturbative unitarity. One can see that a large part of the best-fit region to {\sc Xenon1T} data 
 is ruled out by KLOE.  Nevertheless, the allowed regions still provide a good fit to the Xenon excess. 
As an illustrative example, we indicate with a diamond a benchmark point corresponding to an ALP with a mass of 8 MeV that decays to electrons with a lifetime of about 5 fs and to photons with a branching ratio of order $10^{-5}$ (for $C_{\gamma\gamma} = 0$). The corresponding electron-recoil spectrum at {\sc Xenon1T} is shown in Fig.~\ref{fig:spectrum}. 

Although the region of parameter space with $m_a < 6$ MeV -- top-left in Fig.~\ref{fig:pheno} -- is allowed by  collider constraints, the anomalous magnetic moments generated by the ALP mediator severely constrain this region as we show in the next section.
\begin{figure}
$$\includegraphics[height=0.4\textwidth]{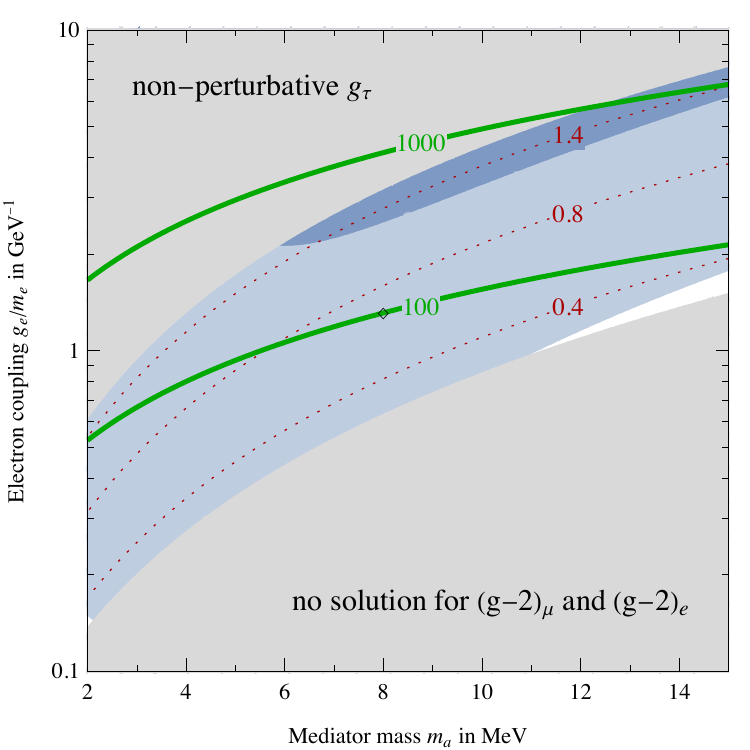} $$
\caption{Parameter space where the central values of $\Delta a_e$ and $\Delta a_{\mu}$ can be explained by additional ALP couplings to leptons. The required tau couplings $g_{\tau}/m_\tau$ are shown as  contour lines  in units of 1/GeV (red, dotted). The contour lines of muon couplings $g_{\mu}/m_\mu$ 
follow those of $g_{\tau}/m_\tau$, and correspond to $g_{\mu}/m_\mu = \{ 3,1,0.7 \} \cdot 10^{-3}$/GeV for $g_{\tau}/m_\tau = \{ 0.4, 0.8, 1.4  \}$/GeV, respectively. Also shown is the  tuning (green, solid) as defined in the text, and the best-fit regions (blue) for reproducing the {\sc Xenon1T} excess, with the diamond denoting the same benchmark point, see Fig.~\ref{fig:pheno}.  \label{plotAeAmu}}
\end{figure}
\section{Constraints from Leptonic Anomalous Magnetic Moments}
The leptonic anomalous magnetic moments (AMMs), $a_\ell = (g-2)_\ell/2$, provide important constraints on light ALPs with couplings to leptons. It is well known~\cite{Chang:2000ii, Marciano:2016yhf, Bauer:2017ris, Bauer:2019gfk, Cornella:2019uxs} that such particles are in fact suitable candidates to simultaneously accommodate the longstanding discrepancy between experimental value and SM prediction for the muon AMM~\cite{Bennett:2006fi, Aoyama:2020ynm, Abi:2021gix}, { $\Delta a_\mu = a_\mu^{\rm exp} - a_\mu^{\rm SM} = (25.1 \pm 5.9) \times 10^{-10}$ (at the level of about $4.2 \sigma$)}.  { The situation regarding the electron AMM~\cite{Hanneke:2008tm, Hanneke:2010au} is inconclusive at the moment, as the theoretical prediction in the SM~\cite{Aoyama:2014sxa} is strongly sensitive to the precise value of the fine structure constant, which has recently been measured with incompatible results. From cesium recoil experiments~\cite{Cs} ones finds $\Delta a_e = (-8.7 \pm 3.6) \times 10^{-13}$, while rubidium experiments give $\Delta a_e = (4.8 \pm 3.0) \times 10^{-13}$~\cite{Rb}. In the following we will use the former (Cs) experimental value, our conclusions will only slightly change if we would have used the more recent Rb value. }

The Lagrangian  in Eq.~\eqref{eq:Lbase2} 
gives a contribution to the AMM of the electron~\cite{Chang:2000ii, Cornella:2019uxs, Bauer:2017ris}
\be
\Delta a_e^{\rm 1loop}  = - \frac{m_e^2}{4 \pi^2 \Lambda^2} \left|c_{ea} \right|^2 h_1\Big( \frac{m_a^2}{m_e^2}\Big) \, ,
\label{1loop}
\ee
where $h_1(x) = \int_0^1 dy\, 2 y^3/(x- x y + y^2)$ is a positive-definite loop function. For the benchmark point in Fig.~\ref{fig:pheno} ($g_e/m_e \sim 1$ GeV$^{-1}$ and $m_a = 8 \, \MeV$), this corresponds to a $\Delta a_e = - 5 \times 10^{-11} (g_e/m_e \GeV) $, which  is about  two orders of magnitudes too large. 

However, allowing for a non-zero coupling to photons $C_{\gamma\gamma}$ in Eq.~\eqref{eq:Lbase2}, there is an additional contribution to $\Delta a_e$
\be\label{eq:deltaagamma}
\Delta a_e^{\gamma\gamma} = -\frac{m_e^2 \alpha}{2\pi^3\Lambda^2}c_{ae} C_{\gamma\gamma} \log{\frac{\Lambda^2}{m_e^2}} + \text{finite terms},\ee
where $\Lambda > m_a$ is a UV scale, and the finite terms can be computed upon specifying a UV completion~\cite{Bauer:2017ris}.
By choosing a coefficient (in the limit $m_a \gg m_e$)
\be\label{eq:cgammatuning}
C_{\gamma\gamma} \approx -c_{ea} \frac{\pi}{\alpha} \frac{m_e^2}{m_a^2} \frac{\log(m_a^2/m_e^2)}{\log(\Lambda^2/m_e^2)}  \, , 
\ee
the photon contribution can cancel the one-loop contribution in Eq.~\eqref{1loop}  to a substantial level, at the price of fine-tuning. 

We now demonstrate that an effective coefficient $C_{\gamma\gamma}$ of the required size can be obtained by introducing additional couplings of $a$ to SM heavy leptons $(\ell = \mu, \tau)$.
In order to do so, it is convenient to work with the Lagrangian in the basis of Eq.~\eqref{eq:Lbase1} setting $\widetilde C_{\gamma\gamma} = 0$. Indeed, this corresponds to $C_{\gamma\gamma} = -c_{\ell a} \approx - c_{e a} $, which up to running effects can be exactly of the right size given in Eq.~\eqref{eq:cgammatuning}. In this basis the $c_{\ell a}$ couplings contribute to the electron AMM via Barr-Zee type diagrams at two-loop order
\be
\Delta a_e^{\rm 2loop}  =  \frac{m_e^2 \alpha}{2 \pi^3 \Lambda^2} c_{ea} c_{\ell a} f \left( \frac{m_a^2}{m_e^2} , \frac{m_a^2}{m_\ell^2}\right) \, , 
\label{2loop}
\ee
where $f(u, v)$ is the loop function 
\begin{align}
f(u, v) = \int_0^1  dx   dy dz \frac{u x   }{u  \overline{x} + u v x y z  \overline{z} + v z  \overline{z} x^2  \overline{y}^2 } \,, 
\end{align}
with the shorthand $ \overline{x} = 1-x$, and similar for $y,z$. When the external lepton mass is small compared to the ALP mass\footnote{In the opposite limit $u,v \ll 1$, $f(u,v) \to 3 - \log v/u.$}, $u \gg 1$, we recover the result in Eq.~(10) of Ref.~\cite{Chang:2000ii}; 
when the lepton mass in the loop is large, $v\ll 1$, we reproduce the effective 1-loop result in Eq.~(37) of Ref.~\cite{Bauer:2017ris}. 
For $u \gg 1$ and $v\ll 1 $, i.e. $m_e \ll m_a \ll m_\ell$, one has $f(u, v) \to 2- \log v$ and $h_1(u) \approx (-11/3 +2 \log u)/u$, so that the two-loop contribution in Eq.~\eqref{2loop} can potentially cancel the one-loop electron contribution in Eq.~\eqref{1loop}, even when $c_{\ell a} \sim c_{ea}$ (see also Ref.~\cite{Alves:2017avw}). 

Therefore one can make the model compatible with the electron AMM by adding a coupling $c_{\tau a}$ of the ALP to tau leptons, which can be tuned to reproduce the central value of $\Delta a_e = - 8.7 \times 10^{-13}$ for the relevant region of parameter space in Fig.~\ref{fig:pheno} { (with a tuning of the same order one could equally obtain the value $\Delta a_e =  4.8 \times 10^{-13}$)}. Moreover, by adding also a coupling $c_{\mu a}$ of the ALP to muons, one can simultaneously account for both $\Delta a_e$ and $\Delta a_\mu$, although only in a subregion of the parameter space. By choosing suitable values $c_{\tau a} \approx c_{e a}$ and $c_{\mu a} \ll c_{e a}$,  $\Delta a_\mu$ is dominated by the 2-loop contribution proportional to $c_{\mu a} c_{\tau a}$. There is also a 
second solution with 
(roughly factor 10) 
larger values for $c_{\mu a}$, but $\Delta a_\mu$ results from a cancellation of 1-loop and 2-loop contributions, leading to an additional tuning. For this reason we focus on the first solution in the following. 

Fig.~\ref{plotAeAmu}  shows the resulting region of parameter space where the central values of $\Delta a_e$ and $\Delta a_{\mu}$ can be explained by additional ALP couplings to heavy leptons. Also shown is the region excluded by perturbative unitarity, the contour lines of $2 c_{\tau a}/\Lambda = g_\tau/ m_\tau$ (red, dotted) and of the required tuning (green, solid). This tuning is defined as $ |\Delta a_e^{\rm 1loop}/ \Delta a_e^{\rm exp}|$ and it is needed to partially cancel the 1-loop contribution to $\Delta a_e$ as  explained above. The contours of $g_\mu/ m_\mu$  follow those of $g_\tau/ m_\tau$, with values indicated in the caption. It is  worth noting that Fig.~\ref{plotAeAmu}  also shows (to very good approximation) the parameter space for the scenario where the muon AMM is not addressed at all, i.e. $c_{\mu a} = 0$, which removes the excluded gray region in the lower right corner.  

\section{Possible UV completions}
Our scenario has  similarities with the ``visible" QCD axion in the MeV range considered in Ref.~\cite{Alves:2017avw}, although we have not considered couplings to quarks. Thus an interesting extension of our model could involve couplings to colored fermions, also enabling a connection to the strong CP Problem. Recently an explicit, phenomenologically viable UV completion of the model in Ref.~\cite{Alves:2017avw} has been proposed in Ref.~\cite{Liu:2021wap} along the lines of classic DFSZ axion models~\cite{Dine:1981rt, Zhitnitsky:1980tq}.  This example demonstrates that it is possible to consider weakly coupled UV-complete models of ALPs at the GeV scale that satisfy all experimental constraints, and provides an explicit (although presumably non-minimal) UV completion to our setup.

On the other hand some ingredients in our scenario rather point  to an UV completion that involve dark strong dynamics. First, the coupling of the mediator to DM must be large; this suggests the possibility that $a$ is the ``pion'' of a dark strong dynamics, with DM being the ``baryon''. Second, the latter is also functional to reproduce the DM relic density as asymmetric DM, since its mass is in the right ball-park and the  p-wave annihilation $\mathrm{DM} \, \mathrm{DM} \to a a $ would efficiently dilute the symmetric component, being larger than the thermal cross-section\footnote{Considering the parameters of the benchmark point in Fig.~\ref{fig:spectrum} we get $\langle \sigma v\rangle \simeq  5 \cdot 10^{-21}$ cm$^3$/s at $x = m_\chi/T = 30$.}.
At the same time, the asymmetric nature of DM would protect from indirect-detection bounds since the s-wave annihilation channel $\mathrm{DM} \, \mathrm{DM} \to ee $ 
is quite large ($\approx  10^{-26}$ cm$^3$/s). 

{Clearly, any realistic UV completion would be subject to further, model-dependent, experimental constraints. Since any of these constraints depends on the specific model considered, we refrain from analyzing them here.}

\section{Discussion and conclusions}
\begin{figure}
$$\includegraphics[width=0.45\textwidth]{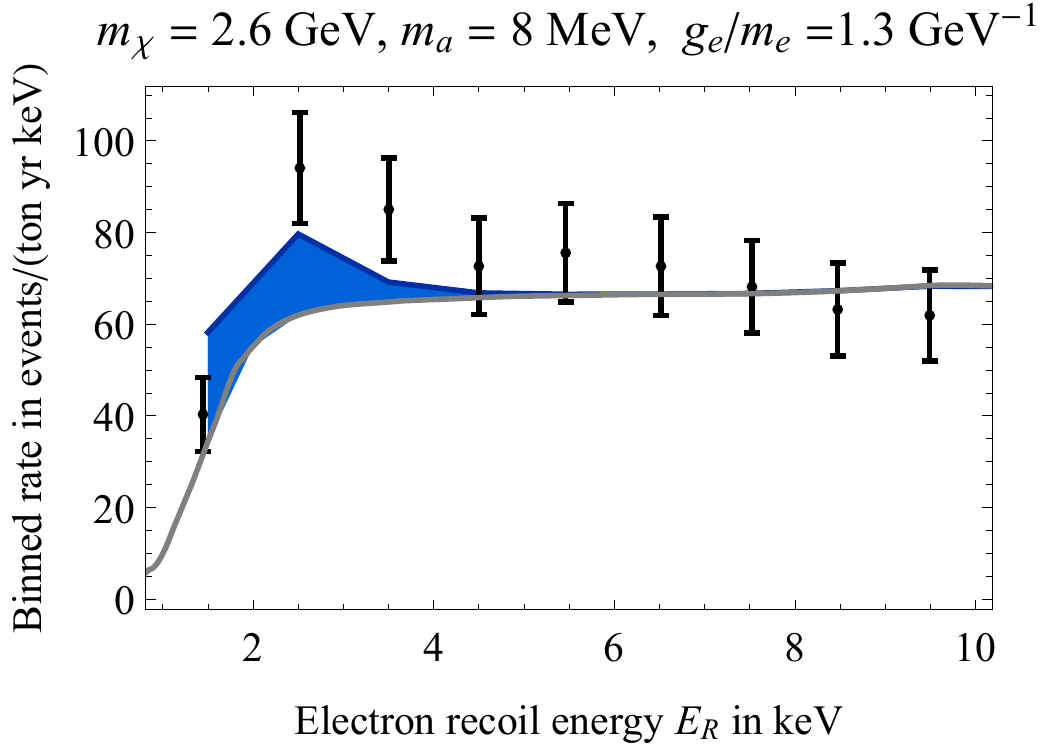}$$
\caption{Example spectrum that fits the excess in \cite{Aprile:2020yad}. The point in the parameter-space shown here is denoted by a diamond in Fig.~\ref{fig:pheno}. 
\label{fig:spectrum}}
\end{figure}

{We have shown that dark-matter-electron scattering mediated by a light pseudo-scalar resonance is able to explain the Xenon1T excess, and account at the same time for the anomalous magnetic moments of muon and electron, at the price of a few percent tuning.}
Our main results are summarized in  Figs.~\ref{fig:pheno} and \ref{plotAeAmu}, which show the experimentally allowed parameter region.
 The quality of the fit of the {\sc Xenon1T} excess is 
 good; even if the region where all constraints are satisfied is 1--2 $\sigma$ away from the model best-fit region (which has $\chi^2/{\rm d.o.f.} \simeq 5.8/7$), the improvement with respect to the Standard Model in explaining the {\sc Xenon1T} data is manifest. This is exemplified in the spectrum shown in Fig.~\ref{fig:spectrum}: the signal in the second and third bins can be explained without over-shooting too much the first one.
{We stress that other mediators, such as vector or scalar bosons, are not able to fit the excess compatibly with all the experimental constraints. Indeed, the non-relativistic suppression of the DM-$e$ pseudo-scalar interaction alleviates the low-recoil constraints (e.g.~the {\sc Xenon} S2-only analysis) that are strong for collisions mediated by a scalar or a vector.}

\smallskip
We note that the experimental {\sc Xenon1T} and {\sc DAMA} recoil spectra are very similar in shape. As a consequence one can be tempted to fit both the anomalies with the  model introduced in this letter. 
However, we have checked that the required cross section to fit {\sc DAMA} is significantly larger than the one needed for {\sc Xenon1T}. 

\smallskip
Finally, we stress that, if the excess will be confirmed by future data, the explanation presented here can be 
 investigated at colliders by searching for the ALP mediator $a$ coupled to electrons, since the allowed parameter region is not far from the existing collider limits.
 Indeed planned experiments such as PADME~\cite{PADME}, VEPP-3~\cite{VEPP3a,VEPP3b} and DarkLight~\cite{DarkLighta, DarkLightb} will probe the entire region of interest. 

\smallskip 

{\small
\section*{Acknowledgements}
We thank Diego Redigolo for discussions and interest in our work. We thank Uli Nierste and Paride Paradisi for useful discussions.  This work is partially supported by project C3b of the DFG-funded Collaborative Research Center TRR 257, ``Particle Physics Phenomenology after the Higgs Discovery". DB is partially supported by MIUR through grant PRIN 2017L5W2PT and by the INFN grant `FLAVOR'. The work of DT was supported in part  by the ERC grant 669668 NEO-NAT.

\clearpage
\newpage


\begin{thebibliography}{99}

\bibitem{Aprile:2020tmw}
E.~Aprile \textit{et al.} [XENON],
``Observation of Excess Electronic Recoil Events in XENON1T''
Phys. Rev. D \textbf{102}, no.7, 072004 (2020)
 [\hhref{arXiv:2006.09721} [hep-ex]].

\bibitem{DiLuzio:2020jjp}
L.~Di Luzio, M.~Fedele, M.~Giannotti, F.~Mescia and E.~Nardi,
``Solar axions cannot explain the XENON1T excess''
Phys. Rev. Lett. \textbf{125}, no.13, 131804 (2020)
 [\hhref{arXiv:2006.12487} [hep-ph]].

\bibitem{Gao:2020wer}
C.~Gao, J.~Liu, L.~T.~Wang, X.~P.~Wang, W.~Xue and Y.~M.~Zhong,
``Re-examining the Solar Axion Explanation for the XENON1T Excess''
Phys. Rev. Lett. \textbf{125}, no.13, 131806 (2020)
[\hhref{arXiv:2006.14598} [hep-ph]].

\bibitem{Kannike:2020agf}
K.~Kannike, M.~Raidal, H.~Veerm\"ae, A.~Strumia and D.~Teresi,
``Dark Matter and the XENON1T electron recoil excess''
\hhref{[arXiv:2006.10735} [hep-ph]].

\bibitem{McKeen:2020vpf}
D.~McKeen, M.~Pospelov and N.~Raj,
``Hydrogen portal to exotic radioactivity''
[\hhref{arXiv:2006.15140} [hep-ph]].

\bibitem{Takahashi:2020bpq}
F.~Takahashi, M.~Yamada and W.~Yin,
``XENON1T anomaly from anomaly-free ALP dark matter and its implications for stellar cooling anomaly''
[\hhref{arXiv:2006.10035} [hep-ph]].

\bibitem{Smirnov:2020zwf}
J.~Smirnov and J.~F.~Beacom,
``Co-SIMP Miracle''
Phys. Rev. Lett. \textbf{125}, no.13, 131301 (2020)
[\hhref{arXiv:2002.04038} [hep-ph]].

\bibitem{Bloch:2020uzh}
I.~M.~Bloch, A.~Caputo, R.~Essig, D.~Redigolo, M.~Sholapurkar and T.~Volansky,
``Exploring New Physics with O(keV) Electron Recoils in Direct Detection Experiments''
[\hhref{arXiv:2006.14521} [hep-ph]].

\bibitem{Cornella:2019uxs}
C.~Cornella, P.~Paradisi and O.~Sumensari,
``Hunting for ALPs with Lepton Flavor Violation''
JHEP \textbf{01}, 158 (2020)
[\hhref{arXiv:1911.06279} [hep-ph]].

\bibitem{Roberts:2019chv}
B.~M.~Roberts and V.~V.~Flambaum,
``Electron-interacting dark matter: Implications from DAMA/LIBRA-phase2 and prospects for liquid xenon detectors and NaI detectors''
Phys. Rev. D \textbf{100}, no.6, 063017 (2019)
[\hhref{arXiv:1904.07127} [hep-ph]].

\bibitem{Roberts:2016xfw}
B.~M.~Roberts, V.~A.~Dzuba, V.~V.~Flambaum, M.~Pospelov and Y.~V.~Stadnik,
``Dark matter scattering on electrons: Accurate calculations of atomic excitations and implications for the DAMA signal''
Phys. Rev. D \textbf{93}, no.11, 115037 (2016)
[\hhref{arXiv:1604.04559} [hep-ph]].

\bibitem{Aprile:2020yad}
E.~Aprile \textit{et al.} [XENON],
``Energy resolution and linearity of XENON1T in the MeV energy range''
Eur. Phys. J. C \textbf{80}, no.8, 785 (2020)
[\hhref{arXiv:2003.03825} [physics.ins-det]].

\bibitem{Aprile:2019xxb}
E.~Aprile \textit{et al.} [XENON],
``Light Dark Matter Search with Ionization Signals in XENON1T''
Phys. Rev. Lett. \textbf{123} (2019) no.25, 251801
[\hhref{arXiv:1907.11485} [hep-ex]].

\bibitem{Alves:2017avw}
D.~S.~M.~Alves and N.~Weiner,
``A viable QCD axion in the MeV mass range''
JHEP \textbf{07}, 092 (2018)
[\hhref{arXiv:1710.03764} [hep-ph]].

\bibitem{E774}
A.~Bross, M.~Crisler, S.~H.~Pordes, J.~Volk, S.~Errede and J.~Wrbanek,
``A Search for Shortlived Particles Produced in an Electron Beam Dump''
Phys. Rev. Lett. \textbf{67} (1991), 2942-2945

\bibitem{E141}
E.~M.~Riordan, M.~W.~Krasny, K.~Lang, P.~De Barbaro, A.~Bodek, S.~Dasu, N.~Varelas, X.~Wang, R.~G.~Arnold and D.~Benton, \textit{et al.}
``A Search for Short Lived Axions in an Electron Beam Dump Experiment''
Phys. Rev. Lett. \textbf{59} (1987), 755

\bibitem{NA64}
D.~Banerjee \textit{et al.} [NA64],
``Improved limits on a hypothetical X(16.7) boson and a dark photon decaying into $e^+e^-$ pairs''
Phys. Rev. D \textbf{101} (2020) no.7, 071101
[\hhref{arXiv:1912.11389} [hep-ex]].


\bibitem{Bjorken:2009mm}
J.~D.~Bjorken, R.~Essig, P.~Schuster and N.~Toro,
``New Fixed-Target Experiments to Search for Dark Gauge Forces''
Phys. Rev. D \textbf{80} (2009), 075018
[\hhref{arXiv:0906.0580} [hep-ph]].


\bibitem{Ema:2020fit}
Y.~Ema, F.~Sala and R.~Sato,
``Dark matter models for the 511 keV galactic line predict keV electron recoils on Earth,''
[arXiv:2007.09105 [hep-ph]].



\bibitem{E137}
J.~D.~Bjorken, S.~Ecklund, W.~R.~Nelson, A.~Abashian, C.~Church, B.~Lu, L.~W.~Mo, T.~A.~Nunamaker and P.~Rassmann,
``Search for Neutral Metastable Penetrating Particles Produced in the SLAC Beam Dump''
Phys. Rev. D \textbf{38} (1988), 3375

\bibitem{Marciano:2016yhf}
W.~J.~Marciano, A.~Masiero, P.~Paradisi and M.~Passera,
``Contributions of axionlike particles to lepton dipole moments''
Phys. Rev. D \textbf{94}, no.11, 115033 (2016)
[\hhref{arXiv:1607.01022} [hep-ph]].

\bibitem{Bauer:2017ris}
M.~Bauer, M.~Neubert and A.~Thamm,
``Collider Probes of Axion-Like Particles''
JHEP \textbf{12}, 044 (2017)
[\hhref{arXiv:1708.00443} [hep-ph]].

\bibitem{Bauer:2019gfk}
M.~Bauer, M.~Neubert, S.~Renner, M.~Schnubel and A.~Thamm,
``Axionlike Particles, Lepton-Flavor Violation, and a New Explanation of $a_\mu$ and $a_e$''
Phys. Rev. Lett. \textbf{124}, no.21, 211803 (2020)
[\hhref{arXiv:1908.00008} [hep-ph]].


\bibitem{Chang:2000ii}
D.~Chang, W.~F.~Chang, C.~H.~Chou and W.~Y.~Keung,
``Large two loop contributions to g-2 from a generic pseudoscalar boson''
Phys. Rev. D \textbf{63}, 091301 (2001)
[\hhref{arXiv:hep-ph/0009292} [hep-ph]].

\bibitem{Bennett:2006fi}
G.~W.~Bennett \textit{et al.} [Muon g-2],
``Final Report of the Muon E821 Anomalous Magnetic Moment Measurement at BNL''
Phys. Rev. D \textbf{73}, 072003 (2006)
[\hhref{arXiv:hep-ex/0602035} [hep-ex]].



\bibitem{Aoyama:2020ynm}
T.~Aoyama, N.~Asmussen, M.~Benayoun, J.~Bijnens, T.~Blum, M.~Bruno, I.~Caprini, C.~M.~Carloni Calame, M.~C\`e and G.~Colangelo, \textit{et al.}
``The anomalous magnetic moment of the muon in the Standard Model,''
Phys. Rept. \textbf{887} (2020), 1-166
[\hhref{arXiv:2006.04822} [hep-ph]].

\bibitem{Abi:2021gix}
B.~Abi \textit{et al.} [Muon g-2],
``Measurement of the Positive Muon Anomalous Magnetic Moment to 0.46~ppm,''
Phys. Rev. Lett. \textbf{126} (2021) no.14, 141801
[\hhref{arXiv:2104.03281} [hep-ex]].


\bibitem{Hanneke:2008tm}
D.~Hanneke, S.~Fogwell and G.~Gabrielse,
``New Measurement of the Electron Magnetic Moment and the Fine Structure Constant''
Phys. Rev. Lett. \textbf{100}, 120801 (2008)
[\hhref{arXiv:0801.1134} [physics.atom-ph]].

\bibitem{Hanneke:2010au}
D.~Hanneke, S.~F.~Hoogerheide and G.~Gabrielse,
``Cavity Control of a Single-Electron Quantum Cyclotron: Measuring the Electron Magnetic Moment''
Phys. Rev. A \textbf{83}, 052122 (2011)
[\hhref{arXiv:1009.4831} [physics.atom-ph]].

\bibitem{Aoyama:2014sxa}
T.~Aoyama, M.~Hayakawa, T.~Kinoshita and M.~Nio,
``Tenth-Order Electron Anomalous Magnetic Moment --- Contribution of Diagrams without Closed Lepton Loops,''
Phys. Rev. D \textbf{91} (2015) no.3, 033006
[erratum: Phys. Rev. D \textbf{96} (2017) no.1, 019901]
[\hhref{arXiv:1412.8284} [hep-ph]].


\bibitem{Cs}
R.~H.~Parker, C.~Yu, W.~Zhong, B.~Estey,  and H.~M\"uller, 
``Measurement of the fine-structure constant as a test of the Standard Model,''
Science {\bf 360}, 191-195 (2018)
[\hhref{arXiv:1812.04130} [physics.atom-ph]].

\bibitem{Rb}
L.~Morel, Z.~Yao, P.~Clad\'e and S.~Guellati-Kh\'elifa, 
``Determination of the fine-structure constant with an accuracy of 81 parts per trillion,''
Nature {\bf 588} no.~7836, 61-65 (2020).


\bibitem{Dine:1981rt}
M.~Dine, W.~Fischler and M.~Srednicki,
``A Simple Solution to the Strong CP Problem with a Harmless Axion''
Phys. Lett. B \textbf{104} (1981), 199-202

\bibitem{Zhitnitsky:1980tq}
A.~R.~Zhitnitsky,
``On Possible Suppression of the Axion Hadron Interactions. (In Russian)''
Sov. J. Nucl. Phys. \textbf{31} (1980), 260
\bibitem{PADME}
M.~Raggi and V.~Kozhuharov,
``Proposal to Search for a Dark Photon in Positron on Target Collisions at DA$\Phi$NE Linac''
Adv. High Energy Phys. \textbf{2014} (2014), 959802
[\hhref{arXiv:1403.3041} [physics.ins-det]].

\bibitem{VEPP3a}
B.~Wojtsekhowski,
``Searching for a U-boson with a positron beam''
AIP Conf. Proc. \textbf{1160} (2009) no.1, 149-154
[\hhref{arXiv:0906.5265} [hep-ex]].

\bibitem{VEPP3b}
B.~Wojtsekhowski, D.~Nikolenko and I.~Rachek,
``Searching for a new force at VEPP-3''
[\hhref{arXiv:1207.5089} [hep-ex]].

\bibitem{DarkLighta}
J.~Balewski, J.~Bernauer, W.~Bertozzi, J.~Bessuille, B.~Buck, R.~Cowan, K.~Dow, C.~Epstein, P.~Fisher and S.~Gilad, \textit{et al.}
``DarkLight: A Search for Dark Forces at the Jefferson Laboratory Free-Electron Laser Facility''
[\hhref{arXiv:1307.4432} [physics.ins-det]].

\bibitem{DarkLightb}
J.~Balewski, J.~Bernauer, J.~Bessuille, R.~Corliss, R.~Cowan, C.~Epstein, P.~Fisher, D.~Hasell, E.~Ihloff and Y.~Kahn, \textit{et al.}
``The DarkLight Experiment: A Precision Search for New Physics at Low Energies''
[\hhref{arXiv:1412.4717} [physics.ins-det]].

\bibitem{Liu:2021wap}
J.~Liu, N.~McGinnis, C.~E.~M.~Wagner and X.~P.~Wang,
``Challenges for a QCD Axion at the 10 MeV Scale,''
[\hhref{arXiv:2102.10118} [hep-ph]].



\end{thebibliography}
\end{document}